# Large-scale solar wind flow around Saturn's nonaxisymmetric magnetosphere

A. H. Sulaiman,[1] X. Jia,[2] N. Achilleos,[3,4] N. Sergis,[5,6] D. A. Gurnett,[1] W. S. Kurth[1]

Corresponding author: A.H. Sulaiman, Department of Physics and Astronomy, University of Iowa, Iowa City, Iowa, USA. (ali-sulaiman@uiowa.edu)

[1]Department of Physics and Astronomy, University of Iowa, Iowa City, Iowa, USA.

[2] Department of Climate and Space Sciences and Engineering, University of Michigan, Ann Arbor, Michigan, USA.

[3]Department of Physics and Astronomy, University College London, London, UK

[4]Centre for Planetary Sciences at UCL/Birkbeck, London, UK

[5]Office of Space Research and Technology, Academy of Athens, Athens, Greece

[6]Institute of Astronomy, Astrophysics, Space Applications and Remote Sensing, National Observatory of Athens, Athens, Greece

## Abstract

The interaction between the solar wind and a magnetosphere is fundamental to the dynamics of a planetary system. Here, we address fundamental questions on the large-scale magnetosheath flow around Saturn using a 3D magnetohydrodynamic (MHD) simulation. We find Saturn's polar-flattened magnetosphere to channel ~20% more flow over the poles than around the flanks at the terminator. Further, we decompose the MHD forces responsible for accelerating the magnetosheath plasma to find the plasma pressure gradient as the dominant driver. This is by virtue of a high-$\beta$ magnetosheath, and in turn, the high-$M_A$ bow shock. Together with long-term magnetosheath data by the Cassini spacecraft, we present evidence of how nonaxisymmetry substantially alters the conditions further downstream at the magnetopause, crucial for understanding solar wind-magnetosphere interactions such as reconnection and shear flow-driven instabilities. We anticipate our results to provide a more accurate insight into the global conditions upstream of Saturn and the outer planets.

## 1. Introduction

Planets possessing an intrinsic dynamo present an obstacle, in the form of a magnetosphere, to the continuous flow of the solar wind. This structure of planetary field lines stands off the interplanetary magnetic field (IMF) embedded in the flow. The size of the dayside magnetosphere is largely controlled by the competing external ram and internal magnetic and plasma pressures between the two regimes, with the combination of different solar wind conditions and internal dynamics resulting in a range of magnetospheric sizes from Mercury to Neptune. The shapes of these planets' magnetospheres are also unique, particularly those of rotationally-dominated Jupiter and Saturn which significantly deviate from axisymmetry about the Sun-planet line. Their internal plasmas, sourced from Io and Enceladus respectively, are equatorially confined into a magnetodisc by the centrifugal forces. As a consequence, the magnetospheres are inflated along the equatorial plane with the poles relatively flattened.

The solar wind is highly supersonic thus forming a detached bow shock upstream of the magnetosphere that decelerates, heats and deflects the flow. The flow in this magnetosheath region is reaccelerated by the action of pressure gradient and magnetic tension forces until the freestream conditions are finally met far downstream. This is an important region of interest for numerous subjects including collisionless shocks, turbulence and solar wind-magnetosphere interactions through reconnection and shear flow-driven instabilities (i.e. Kelvin-Helmholtz). For the latter, flow and magnetic field conditions near the magnetopause are understood to be the primary drivers controlling the exchange of momentum and energy between the solar wind and magnetosphere [*Dungey*, 1961]. One of the earliest works on the magnetosheath predicted the draping pattern of the global magnetic field using an axisymmetric model of the bow shock and magnetopause boundaries [*Spreiter et al.*, 1966; *Spreiter and Stahara*, 1980]. Observations of the draping pattern near Earth's dayside magnetopause were found to be largely consistent with their model [*Fairfield*, 1967; *Crooker et al.*, 1985]. *Slavin et al.* (1985) conducted a comprehensive study on Jupiter and Saturn's interaction with the solar wind where they reported discrepancies between the observed and predicted thicknesses of the magnetosheath. They interpreted this to be a consequence of polar flattening which channels the flow streamlines from longer paths around the flanks to shorter paths over the poles. Further developments were made by *Erkaev et al.* (1996) and *Farrugia et al.* (1998) where a simplified MHD numerical treatment was used to investigate the impact of nonaxisymmetry at Jupiter and Saturn on the solar wind flow by modelling the boundaries as hyperboloids of varying oblateness. They proposed that the IMF exhibits a smooth rotation out of the planet's equatorial plane with this effect greatest at Jupiter (larger degree of polar flattening) and to some extent at Saturn, with a dependence on its upstream orientation.

Since the arrival of Cassini at Saturn, there have been several studies dedicated to the magnetosheath and magnetopause regions underpinning the environment's uniqueness. *Masters et al.* (2012) combined both observations and simulations to reveal that the typical plasma $\beta$ (ratio of plasma to magnetic pressures) conditions, particularly its gradient across the magnetopause, are less favorable for dayside reconnection at Saturn compared to Earth. *Desroche et al.* (2013) adopted the formalism of *Erkaev et al.* (1996) to construct global maps of Saturn's nonaxisymmetric magnetopause highlighting regions where the onsets of large-scale reconnection and the Kelvin-Helmholtz instability are expected under varying IMF conditions. More recently, *Masters* (2015) made further progress by constraining Saturn's dayside reconnection voltage and suggesting its sensitivity to the IMF. Other recent studies used long-term in-situ observations to report the extent of the magnetosphere's polar confinement and dawn-dusk asymmetry [*Pilkington et al.* 2014, 2015]. *Kivelson and Jia* (2014) examined the global magnetosphere simulations in which a rotating pattern of field-aligned currents was introduced to model the magnetospheric periodicity [*Jia et al.*, 2012], and found that the compressional waves launched from the rotating current sources result in a significant dawn-dusk asymmetry of the magnetopause boundary. *Sulaiman et al.* (2014) confirmed, using magnetic field observations with the aid of MHD simulations, that the draping pattern significantly deviates from predictions based on axisymmetric models.

In this paper, we investigate for the first time the three-dimensional large-scale flow pattern of the solar wind around Saturn's magnetosphere not associated with magnetic reconnection. By revisiting magnetic field observations reported by *Sulaiman et al.* (2014) and examining outputs from the Block Adaptive Tree Solar wind Roe-type Upwind Scheme (BATS-R-US) 3D MHD model, we seek to address the following questions: (i) To what extent does nonaxisymmetry have an impact on the distribution of mass flux around such polar-flattened magnetospheres? (ii) What forces accelerate the solar wind at Saturn's magnetosheath? (iii) What are the possible consequences for the conditions near the magnetopause?

## 2. Saturn's Magnetosheath Revisited

The coordinate system used is the planetocentric Cartesian Kronocentric Solar Magnetospheric (KSM) system, with positive $X$ pointing towards the Sun, positive $Y$ orthogonal to the magnetic dipole axis (approximately aligned with the rotation axis at Saturn) and pointing towards dusk, and $Z$ such that the magnetic dipole axis is contained in the $X$-$Z$ plane with positive $Z$ pointing north thus completing the right-hand system. A Saturn radius ($R_S$) is the unit of distance (1 $R_S$ = 60,268 km).

Figure 1 summarizes statistics of the magnetic field and flow properties in Saturn's magnetosheath as revealed by Cassini. Figures 1a and 1b show the distribution of the meridional angle, defined as $\sin^{-1}(B_Z/|\boldsymbol{B}|)$, on the equatorial (X-Y) and meridional (Y-Z) planes respectively [*Sulaiman et al.*, 2014]. Taken together with Figure 1c, the meridional angles in the magnetosheath exhibit a broader (higher standard deviation and kurtosis) distribution in comparison to the IMF, translating to enhancements in $B_Z$. Figure 1d shows the plasma and magnetic pressures in Saturn's magnetosheath with a typical plasma $\beta$ of 10-100 [*Krimigis et al.*, 2004; *Sergis et al.*, 2013] which is substantially higher than at Earth owing to the higher Alfvén Mach number ($M_A$) of Saturn's bow shock as shown in Figure 1e with a median of ~14 [*Sulaiman et al.*, 2016]. We consider plasma pressures only due to the thermal population and not those of very high gyroradii, namely suprathermal and water group ions, since they are not typically associated with the bulk flow as they do not impart a net momentum on one side of the magnetopause.

### 3. Results and Discussion

The 3D global MHD model (BATS-R-US) solves the ideal MHD equations on a 3D Cartesian grid with spacing of 0.5 $R_s$ near the magnetopause and bow shock for this application [*Gombosi et al.*, 2002; *Jia et al.*, 2012]. We set the initial conditions as guided by long-term Cassini observations of the upstream region: $M_A = 14$, $|\mathbf{U}| = 490$ km s$^{-1}$ and IMF [$B_x$, $B_y$, $B_z$] = [0, 0.5, 0] nT. We note that various factors are potentially important in determining the degree of polar flattening, including the mass-loading rate, ionospheric conductance, and the upstream solar wind conditions. Such parameters used in the simulations are all within the nominal ranges of published values. More importantly, the degree of polar flattening is in the range of 15-20% which is consistent with observations by *Pilkington et al.* (2014). As far as this study is concerned, the large scale flow is only affected by the shape of the magnetosphere, regardless of the internal dynamics causing it.

### 3.1 Distribution of Mass Flux

To understand the impact of nonaxisymmetry on the magnetosheath flow, we begin by defining regions of interest in the magnetosheath for comparative purposes. Generally, we will examine the flow asymmetries between the polar and equatorial regions and, in a specific case, the asymmetry between the dawn and dusk sectors within the equatorial region. Figure 2a is a cut at the terminator ($X = 0$ $R_S$) showing the distribution of mass flux as viewed from the Sun. Note the modulus of the mass flux is used to make the anti-sunward flowing magnetosheath mass flux positive. The magnetosheath region is clearly bounded by the cross-sections of the bow shock and the magnetopause: the former being the outer boundary where abrupt changes in fluxes occur (e.g. green to yellow in Figure 2a) and the latter being the inner boundary inside which the fluxes are zero (e.g. yellow to blue in Figure 2a). While the magnetospheric flow field is not a region of interest here, it is retained since ambiguities in precisely identifying the magnetopause boundary make this procedure prone to errors which can have an adverse effect on the magnetosheath color code. The color bar is also capped to enhance the gradients of the displayed property in the magnetosheath.

The effect the nonaxisymmetric magnetosphere has on the flow is immediately obvious in the form of asymmetries in the fluxes. We seek to quantify this flow asymmetry by comparing two cuts from Figure 2a. The first (black) is the northward polar mass flux chosen from the center of Saturn along $+Z$ at $Y = 0$ $R_s$ and the second (red) is the duskward equatorial mass flux chosen from the center of Saturn along $+Y$ at $Z = 0$ $R_s$. Figure 2b superimposes these cuts and the main differences are twofold. First, the bow shock and magnetopause locations are non-aligned with the spacing between the boundaries larger along the pole than the flank. This indicates a thicker polar compared to equatorial magnetosheath. Second, the total mass fluxes, i.e. the areas under the curves between their respective pair of boundaries (MP and BS), differ significantly. We calculate the mass flux to be ~20% greater over the pole than around the flank thereby confirming preferential flow. This estimate of mass flux variation is specifically between the north pole and dusk flank and thus we expect it to be a lower bound since there is an additional asymmetry between the dusk and dawn flanks, which will be discussed below.

### 3.2 Flow Acceleration by MHD Pressures

Figures 3a and 3b are steady-state snapshots of the MHD run displaying speed ($|U|$) at the noon meridian (*X-Z*) and equatorial (*X-Y*) planes respectively. Figure 3b displays the magnetic field lines associated with a duskward IMF, as black arrowed lines. The draping pattern of this typical IMF configuration at Saturn was shown to be in broad agreement with that observed by Cassini [*Sulaiman et al.*, 2014]. Four flow streamlines are highlighted in bold representing flow travelling along both extremes of the polar and equatorial regions. Their upstream segments are displaced by 10 $R_S$ from the subsolar point and are along northward (dashed black), southward (solid black), dawnward (dashed red) and duskward (solid red). The corresponding instantaneous speed profiles along these streamlines are normalized to the solar wind speed of 490 km s$^{-1}$ and are plotted in Figure 3c from far upstream at $X = 40$ $R_s$ to the terminator at $X = 0$ Rs. The bow shock is indicated by the abrupt deceleration at X ~ 22 $R_s$. The streamlines meeting the bow shock were slightly offset from one another due to the asymmetries and were deliberately aligned in the figure for ease of comparison. The general trend among all streamlines is a reacceleration of the flow from just downstream of the bow shock up to the terminator. Although not shown here, it is expected the flow speed must continue to rise, albeit more slowly, until it asymptotically approaches the solar wind speed far downstream.

Two main asymmetries are revealed in Figure 3c: (i) between the polar and equatorial flows with generally the faster flows along northward and southward streamlines compared to the duskward and dawnward streamlines; (ii) between the duskward and dawnward flows with faster flows along dusk. The latter asymmetry is attributed to the rotation of Saturn's magnetopause which can transmit momentum to the incident flow through the generation of a circulation at the viscous-like boundary layer. From Figure 3c, the difference in flow velocities between both flanks is in the range of 35-95 km·s$^{-1}$, which is consistent with recent observations by *Burkholder et al.* (2017) of 75 km·s$^{-1}$. Further, the stagnation point is found to be displaced ~5 % dawnward of the subsolar point which is in close agreement with observations by *Pilkington et al.* (2015) of 7±1 % in the same direction. This can also be noticed later in Figure 4.

From a hydrodynamic perspective, the superposition of a circulation and a freestream flow results in a stagnation point located somewhere on the boundary opposite to the sense of the boundary's rotation. Indeed, this would require the transmittance of momentum from the rotating boundary to the flow through viscous-like boundary effects. The effects of a viscous-like boundary and their role in transferring momentum between the solar wind and magnetosphere through the development of K-H instability has been studied extensively in recent years [*Masters et al.*, 2009; *Delamere et al.*, 2011; 2013; *Ma et al.*, 2015;]. These physical processes at the boundary, however, are beyond the scope of our simulations, which aim to capture the large-scale MHD picture of flow around a nonaxisymmetric body.

The next step is to address the polar versus equatorial asymmetry by examining what forces are responsible for the magnetosheath flow acceleration. The steady-state MHD momentum equation is such that

$$\rho(\boldsymbol{U} \cdot \nabla)\boldsymbol{U} = -\nabla p + \boldsymbol{j} \times \boldsymbol{B} \tag{1}$$

where $\rho$ is the mass density, $\boldsymbol{U}$ the bulk flow velocity, $p$ the scalar plasma pressure, $\boldsymbol{j}$ the current density and $\boldsymbol{B}$ the magnetic field. Equation 1 simply balances the flow-related change of momentum in an arbitrary volume of steady-state plasma with both pressure gradient and electromagnetic forces. The $\boldsymbol{j} \times \boldsymbol{B}$ force comprises both magnetic field gradient and tension. We seek to resolve the contributions from each force on an accelerating plasma volume in the magnetosheath. The streamline highlighted (solid white) in Figure 3a is purely in the magnetosheath and chosen to be midway between the magnetopause and bow shock boundaries. Figure 3d shows the contributions of the MHD forces responsible for the plasma accelerating along that streamline. Note that these accelerations are negative since -$X$ is anti-sunward. To visually improve the profiles in figures 3c and 3d, the data along the streamlines have been smoothed by a five-point running average due to noise induced from numerical calculations of derivatives. This has negligible quantitative effect on the profiles which had a high signal-to-noise ratio to start with. It is clear that the total acceleration, or strictly the velocity gradient, ($\partial \boldsymbol{U}/\partial s$, where $s$ is the path along the chosen streamline) is mainly due to the -$\nabla p$ force compared to the $\boldsymbol{j} \times \boldsymbol{B}$ force. The forces have been normalized by $\rho\boldsymbol{U}$ to yield the velocity gradients from their respective contributions.

The dominance of plasma pressure gradient over electromagnetic forces at Saturn is by virtue of the high plasma $\beta$ (up to two orders of magnitude) which, in turn, is due to the high $M_A$ number bow shock. By taking into account the dominant plasma pressure gradients, the preferential flow over Saturn's poles can be explained purely by streamline paths. Consider a freestream pressure in the solar wind $p_\infty$ which abruptly increases to $p_s$ across the shock ($p_s > p_\infty$). This sets up a pressure gradient between the beginning of the magnetosheath, i.e. immediately behind the bow shock, up to the "end" where $p_\infty$ is recovered far downstream. Thus, to first order, the pressure gradient is

$$-\nabla p \sim \frac{p_s - p_\infty}{L} \qquad (2)$$

where $L$ is the length of the streamline along some path around the planet. Since the paths around the equator are greater than those over the poles, i.e. $L_{equator} > L_{pole}$, the pressure gradients satisfy an inequality such that

$$\frac{p_s - p_\infty}{L_{pole}} > \frac{p_s - p_\infty}{L_{equator}} \qquad (3)$$

This supports the observations by *Slavin et al.* (1985) and confirms their prediction that nonaxisymmetry results in flow being channeled from longer paths around the equator to shorter paths above the flattened poles. This mechanism therefore extends to the other outer planets in the solar system with the effect even more pronounced at planets with a higher degree of nonaxisymmetry such as Jupiter. Additionally, the monotonically increasing $M_A$ of the solar wind with heliocentric distance would translate to higher-$\beta$ downstream of bow shocks and thus we expect pressure gradients to be the dominant drivers of magnetosheath flows around other outer planets. A contrasting scenario to this would be the case study performed by *Lavraud et al.* (2007) where they investigated the passage of a coronal mass ejection (CME) at Earth. The $M_A$ was as low as 2 during this transient, which led to a low-$\beta$ magnetosheath. With the aid of the BATS-R-US global MHD simulations, the authors found the acceleration driven by plasma pressure gradient forces to be as little as 8%, with the dominant acceleration being from both the magnetic gradient and tension forces. Further, they found the acceleration to be asymmetric with preferential flow along the flanks owing to the magnetic forces. Unlike Saturn, however, the aforementioned effects arising from a nonaxisymmetric, rotating magnetosphere exerts on the solar wind flow can be ruled out for the case at Earth.

### 3.3 Implications

Figure 4a is a cut showing the simulated meridional angles of the magnetic field on the *Y-Z* plane at $X = 20$ R$_s$. This region is purely in the magnetosheath, roughly halfway between the boundaries, and contains no magnetospheric regions (see Figures 3a and 3b for visual guidance). While the upstream IMF is purely duskward, it can be seen that there are regions of non-zero $B_Z$ in the magnetic field. However, a clear distinction needs to be made as $B_Z$ can be generated even in an axisymmetric magnetosheath with the same IMF. In such a scenario, magnetic field lines at higher latitudes experience a pull northward from the +*Z*-directed flows above the equator and a pull southward from the –*Z*-directed flows below, while the same field lines remain "hinged" to the -*X*-directed solar wind away from the magnetosheath. The resulting bending of the field lines generates a $B_Z$ component at latitudes above and below the equator. The key finding in Figure 4a is around the subsolar region where a volume of magnetosheath plasma exhibits the most competition between both ±*Z*-directed and ±*Y*-directed flows. The pressure gradients along the poleward directions act to twist the field lines towards the *Z*-direction and are able to overcome the pressure gradients along the *Y*-direction, which act to straighten the field lines back to their original duskward configuration. This net torque is capable of increasing the meridional angle by >10° and we interpret this to be responsible for the much larger spread and kurtosis in $B_z$ observed in the magnetosheath. Note that we have chosen to show a specific cut in the magnetosheath but the twisting continues further downstream and up to the magnetopause. Figure 4c is the same cut showing a distribution of the difference between poleward and equatorial flows. The sectors where poleward flows dominate, i.e. $|U_Z| > |U_Y|$, subtend an angle twice as large as the sectors where equatorial flows dominate, i.e. $|U_Y| > |U_Z|$. Note that the separatrix is not a stagnation line, but a region where $|U_Z|$ and $|U_Y|$ are exactly balanced. As the oblateness of the obstacle increases, e.g. at a more polar-flattened magnetosphere such as Jupiter, the arms of the separatrix will approach each other towards the equator. This would increase the areas of the sectors with dominant poleward flow and conversely decrease those of dominant equatorial flow.

In contrast, Figure 4b shows a special case where the IMF is purely southward. Here the deflection of the magnetic field from the $Z$-axis is signified by $\sin^{-1}(B_Y/|\boldsymbol{B}|)$. Magnetic field lines eastward and westward of the Sun-Saturn line experience a pull from the $+Y$ and $-Y$ directed flows respectively. Contrary to the effect in Figure 4a, the region near the subsolar point shows no deflection since the $Y$-directed flows are unable to twist the field lines by overcoming the $Z$-directed flows which act to straighten them. This particular example is for a lower dynamic pressure than that of Figure 4a, hence the larger area of the magnetosheath cut. Although not shown here, the results remain unchanged for a higher dynamic pressure, provided that $M_A$ is comparable.

Here we exploit long-term Cassini observations of the solar wind, bow shock and magnetosheath with the aid of a 3D MHD simulation to strongly advocate the dominance of plasma pressure gradients to explain preferential flows. It must be mentioned, finally, that this work focuses on a large-scale steady-state picture for the purpose of understanding the fundamentals of solar wind flow around nonaxisymmetric gas giants. The upstream conditions here were typical of those at 10 AU and we have considered the case when the solar wind vector is orthogonal to the planetary dipole (i.e. equinox). Sporadic variations on shorter timescales are possible; for example changes in dynamic pressure have been observed to anti-correlate with the expansion of Saturn's ring current, thus varying the degree of nonaxisymmetry [*Bunce et al.*, 2007; *Sergis et al.*, 2017]. Seasonal variations would change the picture since, for example during southern solstice, the streamlines have shorter paths under the southern hemisphere than over the northern hemisphere and this would introduce a north-south asymmetry. Transient effects, such as the passage of a CME, could considerably lower the $\beta$ in the magnetosheath thus reviving the role of magnetic forces such as the event at Earth examined by *Lavraud et al.* (2007). Regarding spatial effects, works by *Sergis et al.* (2017) and *Kellett et al.* (2011) have highlighted how magnetospheric return flows of hot plasma impose a plasma pressure significant enough to deform the ring current thereby enhancing the dawn-dusk asymmetry. Nevertheless, we do not expect this asymmetry to be as pronounced as the polar-equatorial asymmetry, particularly when addressing averaged large-scale flows. More complicated effects, such as tilt and obliquity at Uranus, would naturally have a consequence and such investigations warrant a future study.

## 4. Conclusions

- Saturn's polar-flattened magnetosphere exhibits preferential flow, channelling mass flux of ~20% more over the poles than around the flanks at the terminator. We expect this at the outer planets, or more generally, any gas giant associated with a rapid planetary rotation in a high-$M_A$ regime. Flow around polar-flattened magnetospheres can be thought of as somewhere between the extreme scenarios of flows around a sphere (streamlines symmetric around all sides) and a wing (streamlines entirely over and below the wing).

- The dominant MHD force responsible for accelerating Saturn's magnetosheath plasma is the plasma pressure gradient compared to the magnetic force. This is by virtue of the high-$\beta$ plasma downstream of a high-$M_A$ bow shock. Plasma pressure gradients become even more important with increasing heliocentric distances where the Mach numbers and consequently $\beta$ are higher, e.g. at Uranus and Neptune.

- The shorter streamline paths over the poles result in a greater pressure gradient. This leads to stronger $Z$-directed compared to $Y$-directed forces and can exert a net torque on magnetic field lines leading to twisting in the $Z$-direction. We interpret this to be most probable cause of enhanced average $B_z$ in the magnetosheath as seen in Cassini observations. The consequence would be conditions at the magnetopause vastly different from those predicted based on assumption of axisymmetry. Whereas the assumption that a northward/southward IMF will likely remain northward/southward at the magnetopause is valid, we argue that the assumption of a duskward/dawnward IMF (typical of the Parker spiral configuration at the outer planets) remaining duskward/dawnward, is not.

## Acknowledgements

Cassini magnetometer data are publicly available via NASA's Planetary Data System. The research at the University of Iowa was supported by NASA through Contract 1415150 with the Jet Propulsion Laboratory.

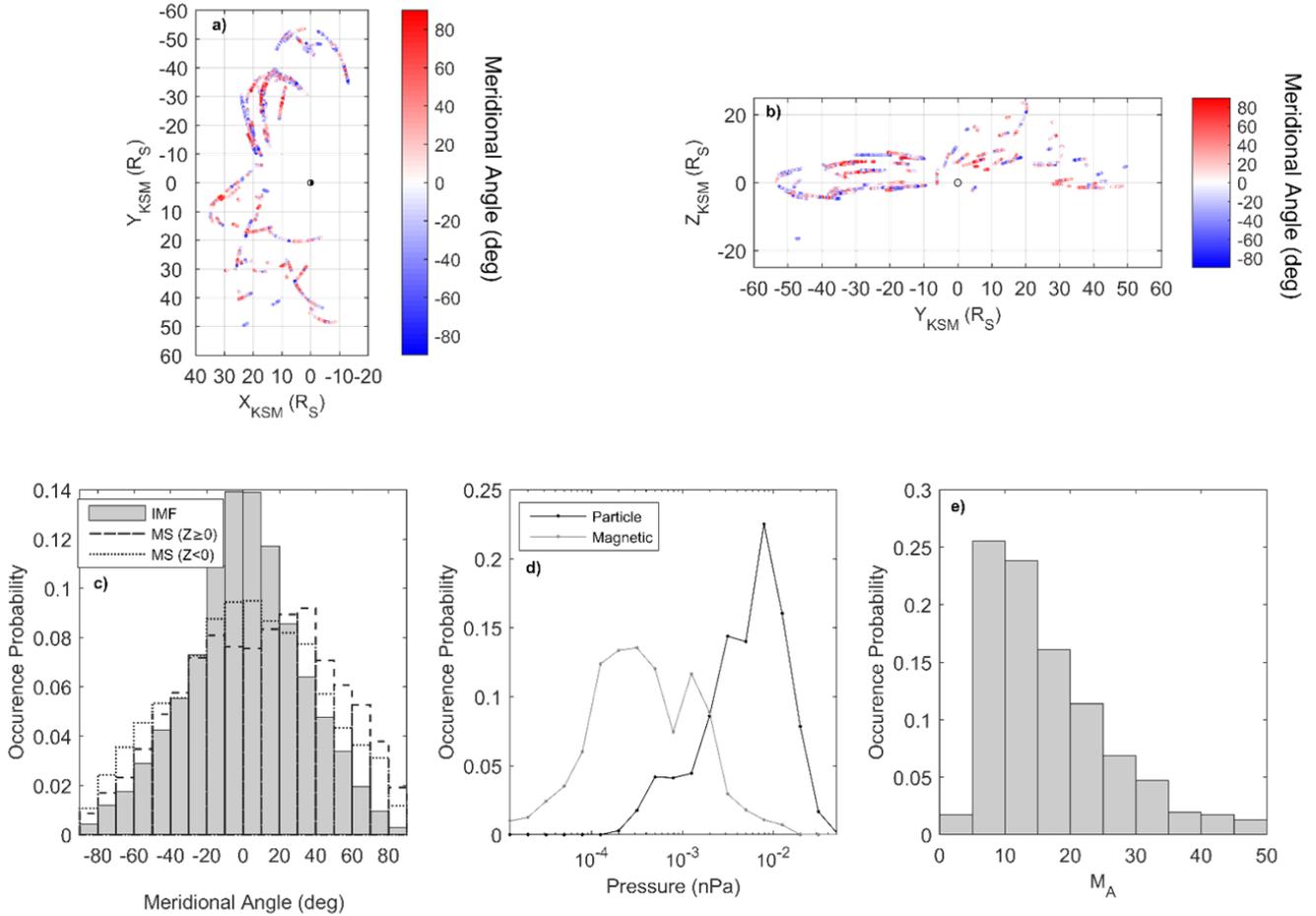

**Figure 1** – Long-term Cassini observations of a) meridional angle, defined as $\sin^{-1} B_Z/|\boldsymbol{B}|$, projected on the equatorial ($X$-$Y$) plane and b) on the meridional ($Y$-$Z$) plane c) distributions of the meridional angles of the IMF (filled bars) and magnetosheath $Z \geq 0$ R$_s$ (dashed bars) and $Z < 0$ R$_s$ (dotted bars) d) distributions of plasma ($< 45$ keV, MIMI instumeny) (black) and magnetic (gray) pressures e) distribution of $M_A$ of Saturn's bow shock

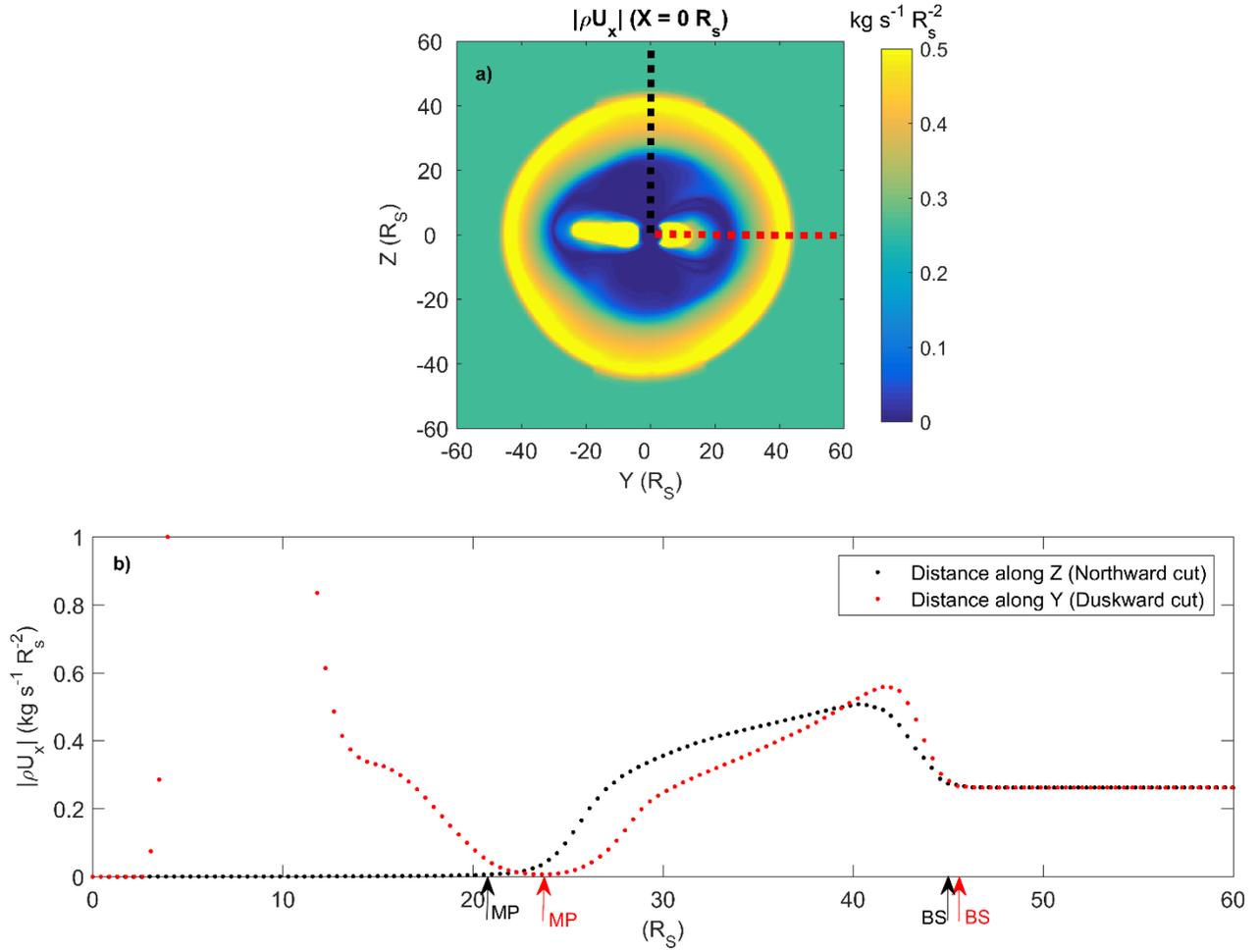

**Figure 2** – a) MHD simulated distribution of the mass flux, as viewed from the Sun (*Y-Z*), at the terminator (*X* = 0 R$_S$). b) Mass fluxes along lines corresponding to (a).

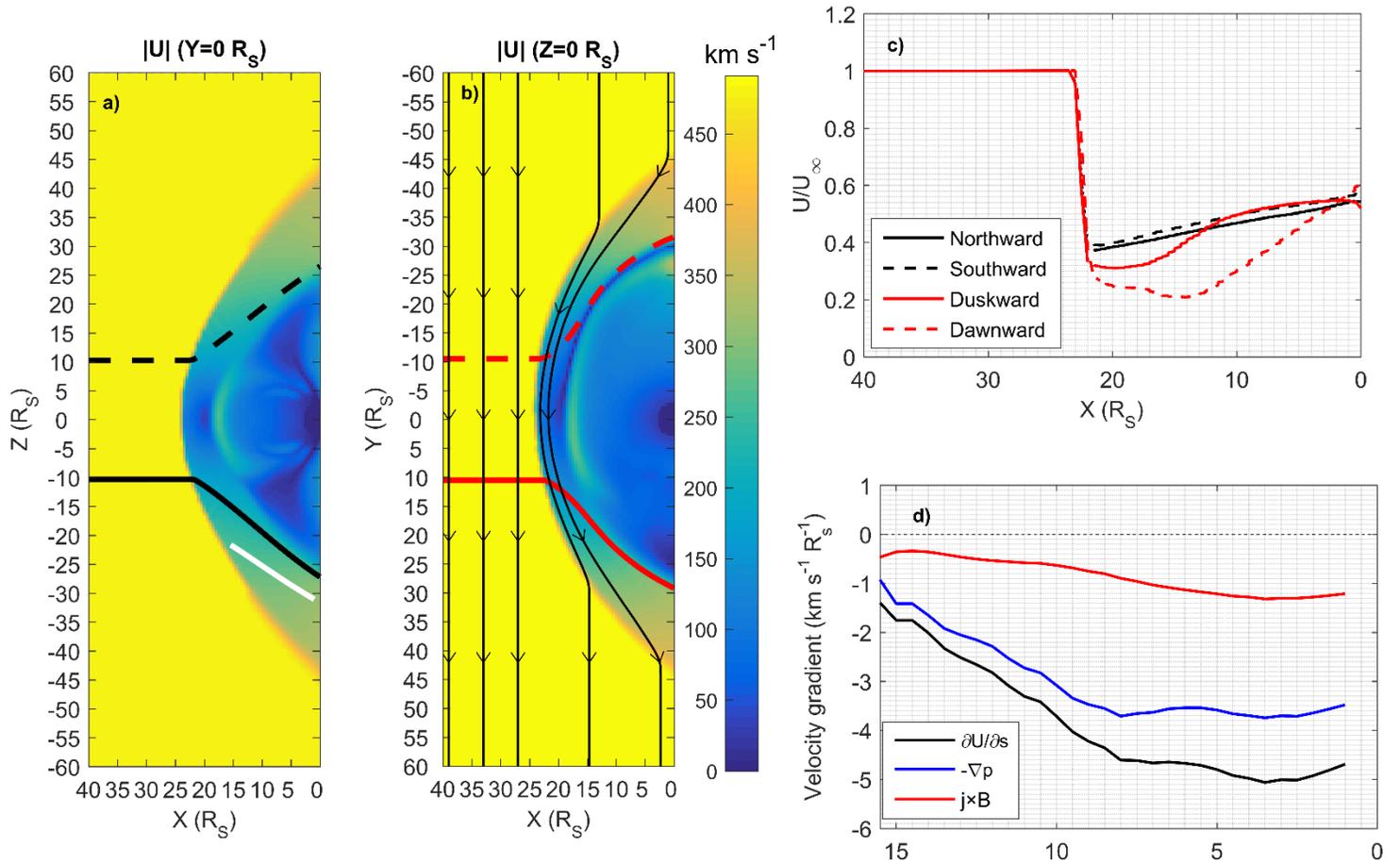

**Figure 3** – MHD simulation cuts of the solar wind speed exterior of the dayside magnetosphere at a) noon meridian (*X-Z*) and b) equatorial (*X-Y*) planes, the latter containing the duskward IMF (black arrows)c) speed profiles (normalized to the solar wind speed of 490 km s⁻¹) plotted along *X* for four streamlines corresponding to those in (a) and (b) representing northward (dashed black), southward (solid black), dawnward (dashed red), and duskward (solid red) flows d) profiles of MHD forces along streamline chosen purely in the magnetosheath (solid white).

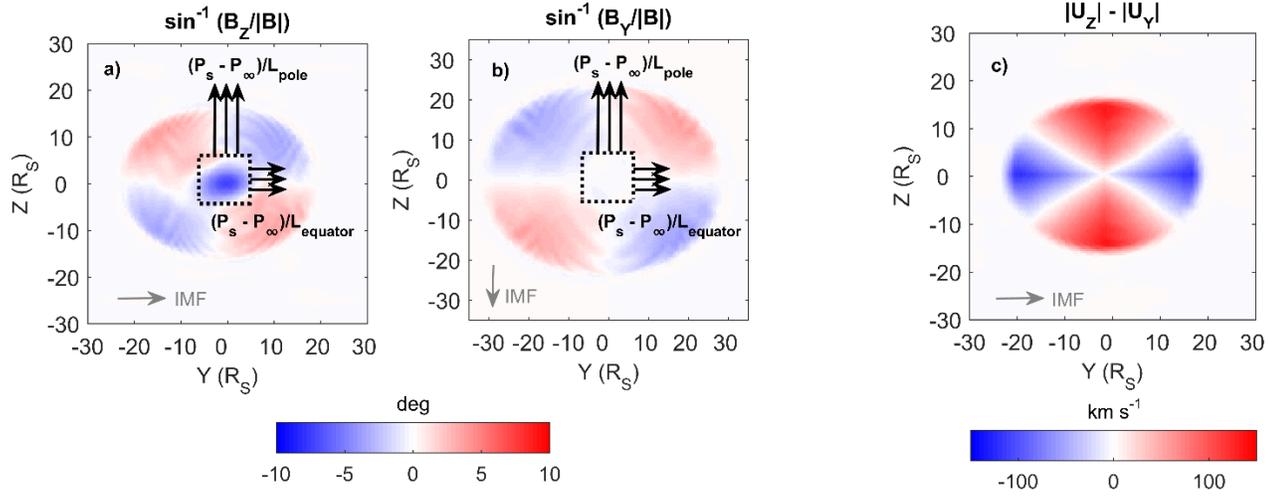

**Figure 4** – Steady-state MHD simulated magnetosheath cuts on the *Y-Z* plane taken close to the magnetopause showing a) distribution of $\sin^{-1}(B_Z/|\boldsymbol{B}|)$ with duskward IMF and b) distribution of $\sin^{-1}(B_Y/|\boldsymbol{B}|)$ with southward IMF. Dotted boxes are the region around the subsolar point where the plasma experiences competing *Z*- and *Y*- directed forces. c) Same magnetosheath cut and setup as a) but with a distribution of $|U_Z| - |U_Y|$. Red regions correspond to $|U_Z| > |U_Y|$ and blue regions $|U_Y| > |U_Z|$.